\documentclass[conference]{IEEEtran}
\IEEEoverridecommandlockouts

\usepackage{cite}
\usepackage{amsmath,amssymb,amsfonts}
\usepackage{algorithmic}
\usepackage{graphicx}
\usepackage{textcomp}
\usepackage{xcolor}
\def\BibTeX{{\rm B\kern-.05em{\sc i\kern-.025em b}\kern-.08em
    T\kern-.1667em\lower.7ex\hbox{E}\kern-.125emX}}

\newcommand{\cS}{{\cal S}}
\newcommand{\cT}{{\cal T}}

\newcommand{\bD}{{\bf D}}
\newcommand{\bE}{{\bf E}}
\newcommand{\bF}{{\bf F}}
\newcommand{\bW}{{\bf \Delta}}
\newcommand{\bfzero}{{\bf 0}}
\newcommand{\bfone}{{\bf 1}}
\newcommand{\bftwo}{{\bf 2}}
\newcommand{\bfthree}{{\bf 3}}
\newcommand{\bffour}{{\bf 4}}

\newcommand{\field}[1]{\mathbb{#1}}

\newcommand{\F}{\field{F}}

\newcommand{\Z}{\field{Z}}

\newtheorem{theorem}{Theorem}
\newtheorem{lemma}{Lemma}
\newtheorem{example}{Example}

\begin{document}

\title{\title{Binary and Non-Binary Self-Dual Sequences\\
and Maximum Period Single-Track Gray Codes}
}

\author{\IEEEauthorblockN{Tuvi Etzion}
\IEEEauthorblockA{\textit{Dept. of Computer Science} \\
\textit{Technion-Israel Institute of Technology}\\
Haifa 3200003, Israel \\
etzion@cs.technion.ac.il}
}

\maketitle

\begin{abstract}
Binary self-dual sequences have been considered and analyzed throughout the years, and
they have been used for various applications.
Motivated by a construction for single-track Gray codes, we examine the structure and recursive constructions
for binary and non-binary self-dual sequences. The feedback shift registers
that generate such sequences are discussed. The connections between these sequences and maximum period single-track codes
are also discussed. Maximum period non-binary single-track Gray codes of length $p^t$ and period $p^{p^t}$ are constructed.
These are the first infinite families of maximum period codes presented in the literature.
\end{abstract}


\section{Introduction}
\label{sec:PM+PR}

A binary self-dual sequence is a cyclic sequence that is invariant under complement.
Such sequences are generated by a feedback shift register known as the complemented cycling register.
They were extensively studied in the fundamental book of shift registers~\cite{Gol67}
and also in a new monograph~\cite{Etz24}. They were also considered, for example, in~\cite{Etz87,Lem70,Wal59}.
These sequences have found many applications, e.g.,
in constructions of de Bruijn sequences with minimal complexity~\cite{BEP96,EtLe84}, in constructions
of single-track Gray codes~\cite{EtPa96}, in constructions of covering sequences~\cite{CETV25},
in constructions of de Bruijn array codes~\cite{Etz25},
and finally in constructions of balanced nearly-perfect covering codes~\cite{BER25}.

Motivated by these applications and mainly by single-track Gray codes, we are trying to find properties of
these sequences, to construct them recursively and efficiently, and to order them for the construction
of single-track Gray codes. The Gray code, which are considered, will be only for maximum period codes.
This will restrict some of the parameters of the self-dual sequences.
We consider the generalization of such sequences over a non-binary alphabet, where
a non-binary self-dual sequence is a sequence that is invariant under the addition of a nonzero constant to
all the elements of the sequence.

The rest of this paper is organized as follows.
In Section~\ref{sec:preli}, the background, necessary definitions, and some known results will be presented.
Section~\ref{sec:binary SD} introduces binary self-dual sequences, especially with the connection to maximum period binary
single-track Gray codes. The discussion will aim to generalize the known results on sequences whose length is a power of two.
Section~\ref{sec:NB_SD} considers the generalization of binary self-dual sequences to non-binary self-dual sequences.
Their recursive construction, properties, and enumeration will be considered.
In Section~\ref{sec:NB-STGC}, the generalization of binary single-track Gray codes to non-binary
single-track Gray codes will be considered. Such codes of maximum period are constructed for length $p^t$ and period $p^{p^t}$,
where $p$ is an odd prime. A short conclusion will be given in Section~\ref{sec:conclusion}.

\section{Preliminaries}
\label{sec:preli}

In this section, we provide the basic definitions and results that are used throughout our exposition.
We also present some background for our work.

A binary cyclic sequence $S = [s_1 s_2 ~ \cdots ~ s_k]$ has {\bf \emph{length}} $k$ if it has $k$ digits.
A binary cyclic sequence $S = [s_1 s_2 ~ \cdots ~ s_k]$ is called a {\bf \emph{self-dual sequence}} (in short, SDS) if it
is equal to its complement. If there is no periodicity in the sequences, then it
can be represented as $S= [X , \bar{X}]$ where $X$ is a binary word of length $k/2$, and $\bar{X}$ is the binary complement of $X$.

\begin{figure}[ht]
\vspace{-0.1cm}
\begin{picture}(90,90)(10,50)
\includegraphics[width=9cm]{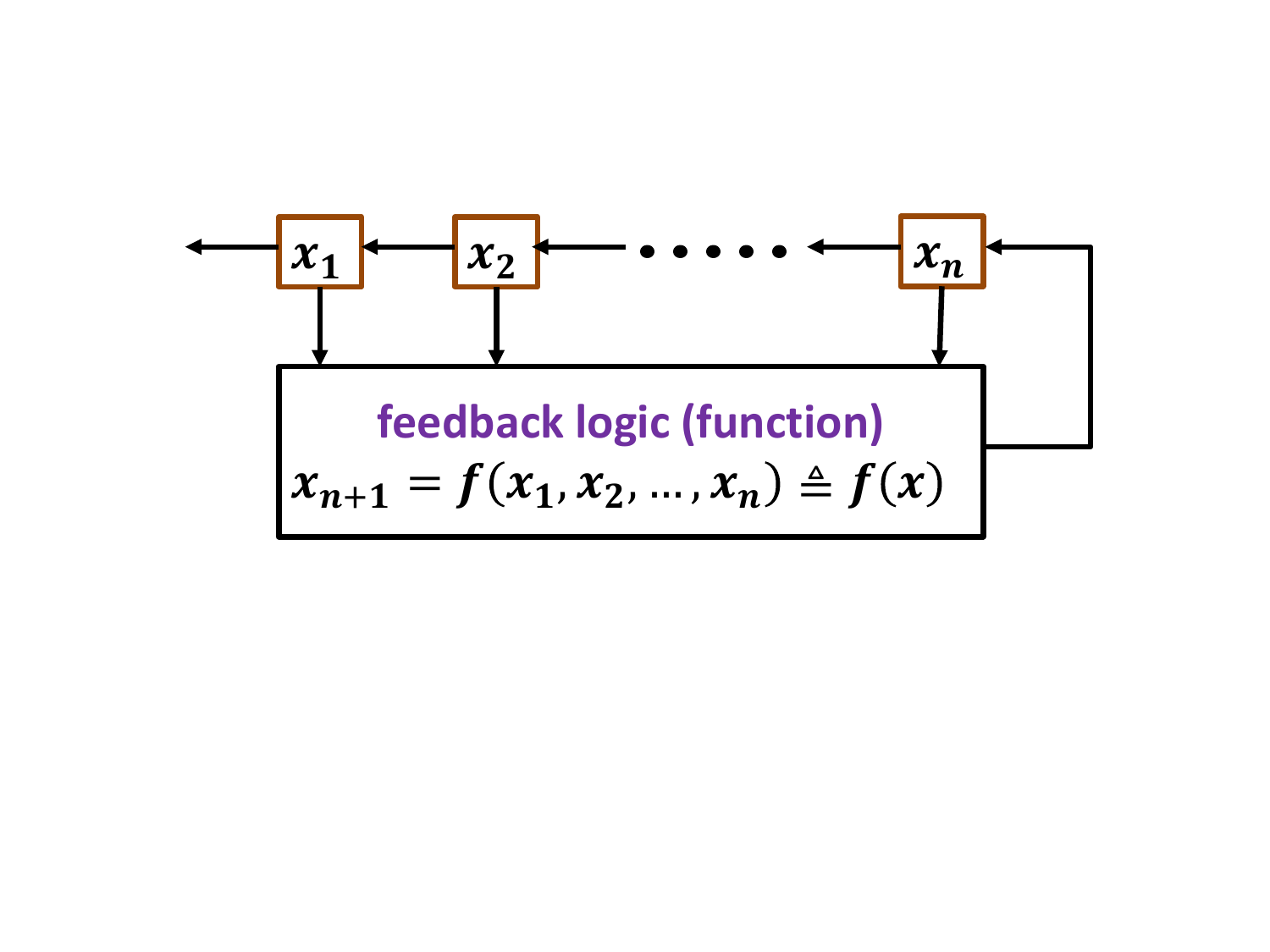}
\end{picture}
\vspace{-1.4cm}
\caption{Feedback shift register of order $n$.}
\label{fig:FSRn}
\end{figure}

A {\bf \emph{feedback shift register}} of order $n$ (in short, an FSR$_n$) has $2^n$ {\bf \emph{states}}, represented by the set
of $2^n$~binary words of length~$n$.
A register has $n$ cells, which are binary storage elements,
where each cell stores at each stage one of the bits of the current state $x=(x_1,x_2,\ldots,x_n)$.
An FSR$_n$ is depicted in Fig.~\ref{fig:FSRn}.

The most relevant FSR$_n$ for our exposition, denoted by CCR$_n$ for the {\bf \emph{complemented cycling register}} of order $n$,
has the feedback function
$$
f(x_1,x_2,\ldots,x_n)=x_1 +1 ~.
$$
The consecutive values obtained from $x_{n+1}$ (until the initial state is repeated),
starting with any initial state is the sequence generated by the register.
The CCR$_n$ generates all the SDSs whose period is a divisor of $2n$ and no other sequences,
where the {\bf \emph{period}} of a sequence $S = [s_1 s_2 ~ \cdots ~ s_k]$ is the length of the shortest word $X$ for which $S=[X , X, \ldots, X]$.
If the FSR$_n$ is linear, then it can also be represented by a polynomial of degree $n$ over the binary field.

The following enumeration was presented in~\cite{Gol67}.
\begin{theorem}
\label{thm:Z*n_Seq}
For each positive integer $n$, the number of sequences generated by the $\textup{CCR}_n$ is
$$
Z^*(n) = \frac{1}{2n} \sum_{\substack{d|n \\ d \textup{~odd}}} \phi(d) \cdot 2^{n/d} .
$$
\end{theorem}

Let SD$(n)$ denote the number of SDSs of period~$n$. A connection between the number of sequences
generated by the CCR$_n$ and the number of SDSs was proved in~\cite{Etz87}.
\begin{lemma}
\label{lem:Z*n_SD}
For each positive integer $n$
$$
Z^*(n) = \sum_{\substack{d|2n \\ d \nmid n}} \textup{SD} (d).
$$
\end{lemma}

Of special interest are SDSs whose length is a power of~2. For these sequences, we have the following interesting phenomena
(see Chapter 5 in~\cite{Etz24}).
\begin{theorem}
All the sequences generated by the polynomial $(1+x)^{2^n +1}$  and not generated by the polynomial $(1+x)^{2^n}$ are all
the SDSs of period $2^{n+1}$. These sequences are exactly the sequences generated by the CCR$_{2^n}$.
\end{theorem}

Two operators play an important role in our exposition. The shift operator $\bE$ is defined on a cyclic sequence as follows:
$$
\bE [s_1 s_2 ~ \cdots ~ s_k] = [s_2 ~ \cdots ~ s_k s_1] ~.
$$

The second operator $\bD$ defined first in~\cite{Lem70} on a cyclic sequence as follows:
$$
\bD [s_1,s_2,\ldots,s_k] = [s_2 - s_1, s_3 - s_2,\ldots, s_k - s_{k-1} ,s_1-s_k ]
$$
$$
=(\bE -\bfone)[s_1,s_2,\ldots,s_k] .
$$

The operator $\bD$ has an inverse operator $\bD^{-1}$. If $S$ is a binary sequence of period $k$ and even weight,
then $\bD^{-1} S$ contains two sequences of period $k$, where one is the complement of the other.
If $S$ is a binary sequence of period $k$ and odd weight, then $\bD^{-1} S$ is an SDS of period $2k$.
The operator $\bD$ and its inverse can be used a few times, i.e.,
$$
\bD^r = \bD (\bD^{r-1}),  ~ \bD^{-r} = \bD^{-1} (\bD^{-r+1}), ~ \text{where} ~ r \geq 2 .
$$
Of special interest is $\bD^{-2^n} S$, where  $S$ is a binary sequence of period $2^{n+1}$.
If $S=[X , Y]$, where $X$ and $Y$ are words of length~$2^n$, then
$$
\bD^{-2^n} S = \bD^{-2^n} [X , Y]
$$
$$
= \{ [ Z , X+Z , X+Z+Y , Z+Y] ~:~ Z \in \F_2^{2^n} \}
$$
(note that plus and minus are the same in the binary field). There are $2^{2^n-1}$ distinct cyclic sequences generated in this way
(if $Z$ is replaced by $Z+X+Y$ the same sequence is obtained). If $Y = \bar{X}$ then
$$
\bD^{-2^n} [X , \bar{X}] = \{ [Z , X+Z , \bar{Z} , X + \bar{Z} ] ~:~ Z \in \F_2^{2^n} \}.
$$
This induces a simple and efficient method to generate all the SDSs of length $2^{n+2}$ from
the SDSs of period $2^{n+1}$. This method was used, for example, in~\cite{EtLe84,EtPa96}.

Finally, we define single-track Gray codes, which are the main motivation for the current exposition.
A length $n$, Period~$P$, {\bf \emph{single-track Gray code}} (in short, STGC) is a list of~$P$ distinct binary words of length~$n$, such that
every two consecutive words, including the last and the first, differ in exactly one position. When
looking at the list as a $P \times n$ array (or an $n \times P$ array),
each column (row, respectively) of the array is a cyclic shift of the first column (row, respectively).
These codes were defined first by Hiltgen, Paterson, and Brandestini~\cite{HPB96}
and considered later in~\cite{EtPa96,FaLi23,QYYS,ScEt99,YLWLLZ,ZZBLZ} for their theoretical value and for their various applications.
Gray codes are subject to extensive research, and an excellent survey on them is presented in~\cite{Mut23}.

It was proved in~\cite{HPB96} that $2n$ must divide $P$, which restricts the possible values of $P$.
A {\bf \emph{maximum period}} STGC is such a code for which $2^n -P <2n$, i.e., there cannot be
such a code of length $n$ and a larger period. To obtain codes with large periods, the two constructions presented
in the following two theorems are used.

\begin{theorem}
\label{thm:single_necklaces}
Let $S_0, S_1,\ldots,S_{r-1}$ be $r$ binary pairwise inequivalent sequences (under cyclic shift) of period $n$,
such that for each $i$, $0 \leq i < r-1$, $S_i$ and $S_{i+1}$ differ in exactly
one coordinate. If there also exists an integer $\ell$, where $\gcd(\ell,n)=1$,
such that  $S_{r-1}$ and $\bE^{\ell} S_0$ differ in exactly one coordinate,
then the following words (read row by row) form a length $n$, period~$nr$ STGC,
\vspace{-0.1cm}
$$
\begin{array}{lclcccl}
S_0 & \hspace{0.3cm} & S_1 & \hspace{0.3cm} & \cdots & \hspace{0.3cm} & S_{r-1} \\
\bE^\ell S_0 & \hspace{0.3cm} & \bE^\ell S_1 & \hspace{0.3cm} & \cdots & \hspace{0.3cm} & \bE^\ell S_{r-1} \\
\bE^{2\ell} S_0 & \hspace{0.3cm} & \bE^{2\ell} S_1 & \hspace{0.3cm} & \cdots & \hspace{0.3cm} & \bE^{2\ell} S_{r-1} \\
\vdots & \hspace{0.3cm} & \vdots  & \hspace{0.3cm} & \vdots & \hspace{0.3cm} & \vdots \\
\bE^{(n-1)\ell} S_0 & \hspace{0.3cm} & \bE^{(n-1)\ell} S_1 & \hspace{0.3cm} & \cdots & \hspace{0.3cm} & \bE^{(n-1)\ell} S_{r-1} \\
\end{array}~.
$$
\end{theorem}

\begin{theorem}
\label{thm:self_dualST}
Let $S_0, S_1,\ldots,S_{r-1}$ be $r$ binary self-dual pairwise inequivalent
sequences of period $2n$. For each $i$, $0 \leq i \leq r-1$, let $S_i =[ s_i^0 , s_i^1,\ldots, s_i^{2n-1}]$ and define
$$
\bF^j S_i = [ s_i^j, s_i^{j+1} ,\ldots, s_i^{j+n-1} ],
$$
where superscripts are taken modulo $2n$.

If for each $0 \leq i < r-1$, $S_i$ and $S_{i+1}$ differ in exactly
two coordinates, and there also exists an integer $\ell$, where $\gcd(\ell,2n)=1$,
such that  $S_{r-1}$ and $\bE^{\ell} S_0$ differ in exactly two coordinates,
then the following words form a length $n$, period $2nr$ STGC,
\vspace{-0.1cm}
$$
\begin{array}{lclcccl}
\bF^0 S_0 & \hspace{0.15cm} & \bF^0 S_1 & \hspace{0.15cm} & \cdots & \hspace{0.15cm} & \bF^0 S_{r-1} \\
\bF^\ell S_0 & \hspace{0.15cm} & \bF^\ell S_1 & \hspace{0.15cm} & \cdots & \hspace{0.15cm} & \bF^\ell S_{r-1} \\
\bF^{2\ell} S_0 & \hspace{0.15cm} & \bF^{2\ell} S_1 & \hspace{0.15cm} & \cdots & \hspace{0.15cm} & \bF^{2\ell} S_{r-1} \\
\vdots & \hspace{0.15cm} & \vdots  & \hspace{0.15cm} & \vdots & \hspace{0.15cm} & \vdots \\
\bF^{(2n-1)\ell} S_0 & \hspace{0.15cm} & \bF^{(2n-1)\ell} S_1 & \hspace{0.15cm} & \cdots & \hspace{0.15cm} & \bF^{(2n-1)\ell} S_{r-1} \\
\end{array} .
$$
\end{theorem}

It was proved in~\cite{ScEt99} that there is no maximum period STGC of length $2^t$ and period $2^{2^t}$, $t >1$.
An STGC of length $2^t$ and period $2^{2^t} - 2^{t+1}$ was constructed in~\cite{EtPa96}.
Non-binary STGCs of length $p^m$ and period $p^{p^m}$ are discussed in Section~\ref{sec:NB-STGC}.
These are the only possible maximum period STGCs that contain all the words of the given length.
For each odd prime $p$, we might have a maximum period STGC of length $p$ and period $2^p-2$.
Such codes are known to exist for all primes up to $p=19$~\cite{Etz07}, and they are constructed
using the construction of Theorem~\ref{thm:single_necklaces}, but they can also be obtained by the construction of Theorem~\ref{thm:self_dualST}.
Maximum period STGC using these theorems can also be obtained for $n=9$ and $n=25$.
For other values of $n$, the construction of Theorem~\ref{thm:single_necklaces} cannot yield a maximum period STGCs.
The situation for the construction of Theorem~\ref{thm:self_dualST} is quite different.
If $n=2p$, then there is exactly one SDS of period 4 and $\frac{2^{2p}-4}{4p}$ SDSs of period $4p$ generated
by the CCR$_{2p}$. It might be possible, in this case, to use the construction of Theorem~\ref{thm:self_dualST} to generate
a maximum period STGC. If $n=4p$, then there are exactly two SDSs of period 8 and $\frac{2^{4p}-16}{8p}$ SDSs of period $8p$ generated
by the CCR$_{4p}$. It might be possible, in this case, to use the construction of Theorem~\ref{thm:self_dualST} to generate
a maximum period STGC. Such analysis can be further given.
If $n=8p$, then there are exactly sixteen SDSs of period 16 and $\frac{2^{8p}-256}{16p}$ SDSs of period $16p$ generated
by the CCR$_{8p}$ and it might be possible to use the construction of Theorem~\ref{thm:self_dualST} to generate
a maximum period STGC when $p >16$. The same process can be analyzed for $n=2^i p$, $i \geq 1$.

\section{Binary Self-Dual Sequences}
\label{sec:binary SD}

The operation $\bD^{2^n}= (\bE+1)^{2^n}$ and its inverse are associated with the sequences generated by the polynomials
$(x+1)^k$. They are generalized for binary sequences whose length is not a power of 2 in the following way.
Let $\bW_m$ be the following operator defined on binary sequences whose length is $r m$.
Let $S = [ X_1 ,X_2 , \ldots , X_r ]$ be a binary sequence, where $X_i$, $1 \leq i \leq r$, is a word of length of length $m$.
The operator $\bW_m$ is defined on $S$ as
\vspace{-0.22cm}
$$
\bW_m S = [ X_1 + X_2 , X_2 + X_3 , \ldots , X_r + X_1 ]
$$
and it inverse $\bW_m ^{-1}$ is defined on $S$ as
\vspace{-0.35cm}
$$
\bW_m ^{-1} S = \{ [Y , Y + X_1 , Y + X_1 + X_2 , \ldots , Y + \sum_{i=1}^{r-1} X_i ] : Y \in \F_2^m \}
$$
if $\sum_{i=1}^r X_i $ is the all-zero word, where $Y$ can be any binary word of length $m$.
If $\sum_{i=1}^r X_i = Z \neq \bfzero$ (the all-zero word), then
\vspace{-0.3cm}
$$
\bW_m ^{-1} S = \{[Y , Y + X_1 , Y + X_1 + X_2 , \ldots , Y + \sum_{i=1}^{r-1} X_i ,Y + Z, 
$$
\vspace{-0.3cm}
$$
Y+Z+ X_1 , Y+Z+X_1 + X_2, \ldots , Y +Z+ \sum_{i=1}^{r-1} X_i] ~:~ Y \in \F_2^m \}.
$$

If $\cS$ is a set of sequences, then we define $\bW_m \cS$ to be the set $\{ \bW_m T ~:~ T \in \cS \}$.
and $\bW_m^{-1} \cS$ is the set $\{ \bW_m^{-1} T ~:~ T \in \cS \}$. It is readily verified that
for any sequence $S$ of length $rm$ we have that $S=\bW_m (\bW_m^{-1} S)$ and $S \in \bW^{-1}_m (\bW_m S)$.

\begin{theorem}
\label{thm:rec_CCR}
If $\cS$ is the set of SDSs generated by the CCR$_n$, then $\bW_n^{-1} \cS$ is the set
of SDSs generated by the CCR$_{2n}$.
\end{theorem}
\begin{IEEEproof}
An SDS of length $4n$ (might be of a smaller period) generated by the CCR$_{2n}$ can be written in the form
$[X , Y , \bar{X} , \bar{Y} ]$, where the length of $X$ and $Y$ is $n$. Hence
$\bW_n [X , Y , \bar{X} , \bar{Y} ]= [X+Y , Y+\bar{X}, \bar{X} +\bar{Y}, \bar{Y} +X]=[X+Y , Y+\bar{X}]$ has length $2n$ and it is
an SDS generated by the CCR$_n$.

An SDS of length $2n$ generated by the CCR$_n$ can be written in the form
$[X , \bar{X}]$, where the length of $X$ is $n$. Hence, $\bW_n^{-1} [X , \bar{X}]= \{[Y , X+Y, \bar{Y} , X + \bar{Y}] ~:~ Y \in \F_2^n \}$
has SDSs of length $4n$ and each one is an SDS generated by the CCR$_{2n}$.
\end{IEEEproof}

\begin{example}
\label{ex:p=3}
There are two SDSs generated by the CCR$_3$, $[000111]$ and $[01]$. By applying $\bW_3^{-1}$ on these two sequences, the six
sequences of the CCR$_6$ are obtained. Four sequences
\vspace{-0.2cm}
$$
\begin{array}{lcc}
& [000 ~~ 000 ~~ 111 ~~ 111] & [001 ~~ 001 ~~ 110 ~~ 110] \\
& [010 ~~ 010 ~~ 101 ~~ 101] & [011 ~~ 011 ~~ 100 ~~ 100] \\
\end{array}
\vspace{-0.2cm}
$$
of period 12
are obtained by applying $\bW_3^{-1}$ on $[000111]$.
One sequence of period 12, $[000 ~~ 010 ~~ 111 ~~ 101]$, and one sequence of period~4, $[0011]$, are
obtained by applying $\bW_3^{-1}$ on $[01]$.

We can continue and apply $\bW_6^{-1}$ on these 6 SDSs of the CCR$_6$. By applying $\bW_6^{-1}$
on the five SDSs of period 12, we obtain 160 SDSs of period 24, which are generated by the CCR$_{12}$.
By applying $\bW_6^{-1}$ on the sequence $[0011]$ we obtain the two SDSs $[00001111]$ and $[00101101]$ of period~8
and ten SDSs of period 24, all generated by the CCR$_{12}$. \hfill\quad $\blacksquare $
\end{example}


The results obtained in Example~\ref{ex:p=3} can be proved in general.
\begin{theorem}
\label{thm:enum_CCR}
The CCR$_n$, $n=2^i p$, generates $2^{2^i-i-1}$ SDSs of period $2^{i+1}$ and $\frac{2^{2^i p} -2^{2^i}}{2^{i+1}p}$
SDSs of period $2^{i+1}p$.
\end{theorem}

The recurrence of Theorem~\ref{thm:rec_CCR} and the enumeration of Theorem~\ref{thm:enum_CCR} and Example~\ref{ex:p=3}
can be further generalized and analyzed. From the $\frac{2^{2^{i+1} p} -2^{2^{i+1}}}{2^{i+2}p}$ SDSs of period $2^{i+2}p$
only $\frac{2^{2^i(p+1)}-2^{2^{i+1}}}{2^{i+2}p}$ are generated from SDSs of period $2^{i+1}$, where each such SDS
of period $2^{i+2}p$ can be generated using any one of $p$ different words of length $2^i p -1$. It can be analyzed which words of length $2^i p -1$
yield SDSs of period $2^{i+2}$ and which yield SDSs of period $2^{i+2}p$.

This analysis of these SDSs is analogous to the analysis of SDSs whose period is a power of 2,
which led to the construction of STGC of length $2^t$ and period $2^{2^t} -2^{t+1}$ is~\cite{EtPa96}.
Another recursive method to generate the SDSs is by interleaving SDSs.
This important method that also apply for non-binary sequences will be analyzed in the full version of this paper.

\vspace{-0.1cm}

\section{Non-binary Self-Dual Sequences}
\label{sec:NB_SD}

\vspace{-0.1cm}

Linear recurring sequences over rings were extensively studied, but the associated shift registers were
hardly investigated (see~\cite{KKMN,ReSl85} and related references) and the same is for non-binary self-dual sequences.
The definition of the CCR$_n$ is genralized for an FSR$_n$ over $\Z_m$.
The $m$-CCR$_n$ is an FSR$_n$ over $\Z_m$ with feedback function
\vspace{-0.15cm}
$$
f(x_1,x_2,\ldots,x_n)=x_1 +1 ~,
$$
where $x_i \in \Z_m$, $1 \leq i \leq n$, and the computation is performed modulo $m$.
The function can also be taken as $f(x_1,x_2,\ldots,x_n)=x_1 +r$, where $r$ is relatively prime to $m$.
A non-binary SDS over $\Z_m$ is a cyclic sequence which is left invariant when a \emph{one} is add to each of its coordinates.
The enumerations of the number of sequences of the CCR$_n$ (see Theorem~\ref{thm:Z*n_Seq} and Lemma~\ref{lem:Z*n_SD}) can be generalized.

%
%
%

Let $[X ~~ X+\bfone ~~ X+\bftwo]$ be an SDS of period $3^n$ over $\Z_3$, where $X$ is a word of length $3^{n-1}$ over $\Z_3$.
Let $Z$ be a word of length $3^{n-1}$ over $\Z_3$ that starts with \emph{zero} and let $Y$ be a word of length $3^{n-1}$ over $\Z_3$.
\begin{lemma}
\label{lem:SDS_Z3}
The sequence $[V , V+\bfone , V+\bftwo]$ (a bold integer is a sequence with only this integer), where $V=(Z , Z+Y , Z+2Y+X)$,
is an SDS of period $3^{n+1}$ over $\Z_3$. Each different choice of $Z$ and $Y$ yields a distinct SDS.
Each SDS of period $3^{n+1}$ over $\Z_3$ is constructed in this way.
\end{lemma}
\begin{IEEEproof}
The proof has similarity to that of Theorem~\ref{thm:rec_CCR} with one significant exception.
By the definition of $\bD^{-1}$ we have
\vspace{-0.17cm}
$$
\bD^{-3^{n-1}} [X , X+\bfone , X+\bftwo]
$$
$$
= \{[Y,Y+X, Y+2X+ \bfone] ~:~ Y\in \F_3^{3^{n-1}} \},
$$
\vspace{-0.4cm}
$$
\bD^{-3^{n-1}} [Y,Y+X, Y+2X+ \bfone]
$$
$$
= \{[V , V+\bfone , V+\bftwo] ~:~ V=Z , Z+Y , Z+2Y+X, ~ Z \in n \F_3^{3^{n-1}} \}
$$
(note that $(\bE-1)^{2^k} = \bE^{2^k}-1$).
Since either $V$ or $V+\bfone$ or $V+\bftwo$ starts with a \emph{zero}, it was sufficient to
consider for $Z$ only those words of length $3^{n-1}$ that start with a \emph{zero}.

Clearly, each sequence $S=[V , V+\bfone , V+\bftwo]$ is an SDS and for each sequence of period $3^{n+1}$ of this form we have that
\vspace{-0.14cm}
$$
\bD^{3^{n-1}} (\bD^{3^{n-1}} S) = [X , X+\bfone , X+\bftwo].
$$
The sequence $[X , X+\bfone , X+\bftwo]$ is an SDS of period $3^n$.

This implies that each SDS of length $3^{n+1}$ is constructed in this way. Now, a simple enumeration shows that
each different choice of $Z$ and $Y$ yields a distinct SDS.
\end{IEEEproof}

The same technique and expression presented in Lemma~\ref{lem:SDS_Z3} for $\Z_3$, can be applied for
any alphabet $\Z_m$, where $m$ is a prime. For example, if $m=5$, then we start with a
SDS Let $S=[X , X+\bfone , X+\bftwo , X+\bfthree , X+\bffour]$ over $\Z_5$, where $X$ is a word of length $5^{n-1}$.
Let $Y_1$, $Y_2$, and $Y_3$ be words of length $5^{n-1}$ and let $Z$ be a word of length $5^{n-1}$ that starts with a \emph{zero}.
From $S$, $Y_1$, $Y_2$. $Y_3$, and $Z$, we form the SDS
$[V , V+\bfone , V+\bftwo , V+\bfthree , V+\bffour]$, where
$$
V = (Z , Z+Y_3 , Z+2Y_3 + Y_2 ,
$$
\vspace{-0.4cm}
\begin{equation}
\label{eq:SD_binom}
Z + 3Y_3 +3Y_2 + Y_1 , Z+4Y_3 + Y_2 + 4Y_1 + X)
\end{equation}
is a word of length $5^n$ over $\Z_5$. Generally, when an SDS over a prime alphabet $\Z_p$ is constructed recursively,
these expressions in the construction, as the one
presented in~(\ref{eq:SD_binom}), are associated with the top $p$ rows of Pascal's triangle, where the entries of the triangle are taken
modulo~$p$. The exact expression will be proved in the full version of the paper.

\section{Non-Binary Single-Track Gray Codes}
\label{sec:NB-STGC}

In this section, we present and discuss the ordering of non-binary SDSs for the construction of maximum period non-binary STGCs.
The requirement is that every two consecutive codewords of length $n$ (not SDSs) will differ in exactly one coordinate.
If it was also required that the difference in this coordinate between the second
and the first of these two consecutive codewords is a given constant $\delta$, then it was proved in~\cite{ScEt99} that
such an STGC exists only for binary codewords of length 2.
Theorem~\ref{thm:single_necklaces} is readily generalized for sequences over $\Z_m$ without any modification.
Theorem~\ref{thm:self_dualST} is generalized to $\Z_m$ with a simple modification as follows.

\begin{theorem}
\label{thm:self_dualST_NB}
Let $S_0, S_1,\ldots,S_{r-1}$ be $r$ binary self-dual pairwise inequivalent full-order SDSs
of length $mn$ over $\Z_m$.
If for each $i$, $0 \leq i < r-1$, $S_i$ and $S_{i+1}$ differ in exactly
$m$ coordinates, and there also exists an integer $\ell$, where $\gcd(\ell,mn)=1$,
such that  $S_{r-1}$ and $\bE^{\ell} S_0$ differ in exactly $m$ coordinates,
then the following words form a length $n$, period $mnr$ STGC,
$$
\begin{array}{lclcccl}
\bF^0 S_0 & \hspace{0.08cm} & \bF^0 S_1 & \hspace{0.08cm} & \cdots & \hspace{0.08cm} & \bF^0 S_{r-1} \\
\bF^\ell S_0 & \hspace{0.08cm} & \bF^\ell S_1 & \hspace{0.08cm} & \cdots & \hspace{0.08cm} & \bF^\ell S_{r-1} \\
\bF^{2\ell} S_0 & \hspace{0.08cm} & \bF^{2\ell} S_1 & \hspace{0.08cm} & \cdots & \hspace{0.08cm} & \bF^{2\ell} S_{r-1} \\
\vdots & \hspace{0.08cm} & \vdots  & \hspace{0.08cm} & \vdots & \hspace{0.08cm} & \vdots \\
\bF^{(mn-1)\ell} S_0 & \hspace{0.08cm} & \bF^{(mn-1)\ell} S_1 & \hspace{0.08cm} & \cdots & \hspace{0.08cm} & \bF^{(mn-1)\ell} S_{r-1} \\
\end{array}.
$$
\end{theorem}

\begin{example}
For $m=3$ and $n=3$, there are three SDSs of length 9 which contain each
word of length three over $\Z_3$ exactly once. These three sequences are the first three columns of the following array.
The other 27 columns of the first three rows form an $3 \times 27$ array, which is a length 3 period 27 STGC over $\Z_3$.
Each column of the $9 \times 27$ array after the first three columns is the associated SDS in the required shift.
Note that also each of the rows of this $9 \times 27$ array is an SDS.
\vspace{-0.1cm}
$$
\begin{array}{lcc}
 &  001 & 001122222~~112200000~~220011111 \\
 &  011 & 011111001~~122222112~~200000220 \\
 &  000 & 000220011~~111001122~~222112200 \\
 \hline
 &  112 & 112200000~~220011111~~001122222 \\
 &  122 & 122222112~~200000220~~011111001 \\
 &  111 & 111001122~~222112200~~000220011 \\
 \hline
 &  220 & 220011111~~001122222~~112200000 \\
 &  200 & 200000220~~011111001~~122222112 \\
 &  222 & 222112200~~000220011~~111001122 \\
\end{array}.
$$
\end{example}

We consider now a construction and an arrangement of the $m^{m-2}$ SDSs of length $m^2$ over $\Z_m$,
which is an easier way to form an STGC of length $m$ and period $m^m$.
Such an arrangement with some more requirement will be used as a seed for the recursive construction
for an STGC of length $p^t$ with period $p^{p^t}$, where $p$ is an odd prime.
We have to order the SDSs of length $m^2$ lexicographically such that the first $m$ digits will start with a \emph{zero}
and the sum of these digits will be 0~modulo~$m$. This simple ordering contains $m^{m-2}$ sequences of length~$m^2$. Each word
of length~$m$ over~$\Z_m$ is a subsequence of exactly one of these sequences. Now, we use Theorem~\ref{thm:self_dualST_NB}
to form an STGC.

But there might be a simpler way to form such an ordering.
The first $m$ digits (and clearly $m+1$ digits) determine the whole SDS. Let $\cT$ be the ordered list of these $m+1$ digits in the STGC.
Assume that the first such $m+1$ digits in a word of $\cT$ are $0x_2 x_3 ~ \ldots ~ x_m 1$.
Consider now the differences $x_2,x_3 - x_2 , x_4 -x_3,\ldots, x_m - x_{m-1}, 1 - x_m$ in this word.
Each word of length $m+1$ in $\cT$ defines a unique such sequence of differences.
The sum of these differences is 1 modulo $m$.

We order this set of differences lexicographically.
In this order, there are $m^{m-1}$ sequences of differences of length~$m$, where the sum of the differences in each sequence is 1~modulo~$m$.
Two sequences of differences are associated with the same SDS if and only if one is a cyclic shift of the other.
From all possible $m$ cyclic shifts, only one sequence of differences is taken.
For each sequence of differences, there are $m$ distinct cyclic shifts since
the sum 1~modulo~$m$ of the differences implies that there is no periodicity in a sequence of differences.

Therefore, there are $m^{m-2}$ cyclic sequences of differences in the final set of differences.
This set of sequences containing the differences is ordered in a way that every two consecutive sequences (including the last and the first)
have an identical subsequence of length $m-2$. Therefore, in the corresponding SDSs of length $m^2$,
the $m^2$~subsequences of length $m$ can be paired such that each pair differs in exactly one coordinate.

\begin{example}
For $n=m=4$, the 16 sequences of differences are ordered in a $4 \times 16$ array such that every
two consecutive columns have an identical subsequence of length~2.
\vspace{-0.09cm}
$$
\begin{array}{lc}
 &  0012232331111100 \\
 &  1111001120021123 \\
 &  0222333322130332 \\
 &  0210033222333000 \\
\end{array}.
$$
Each sequence of differences is translated to a word of length~5, which is extended to an SDS of length~16.
These SDSs are the columns of the following $16 \times 16$ array. The top four rows in this array form
the arrangement of the 16 SDSs to satisfy the conditions of Theorem~\ref{thm:self_dualST_NB}.
This arrangement leads to an STGC of length 4 and period 256.
\vspace{-0.09cm}
$$
\begin{array}{lc}
 &  0032211000000000 \\
 &  0000003331111100 \\
 &  1111000011132223 \\
 &  1333333333222111 \\
 \hline
 &  1103322111111111 \\
 &  1111110002222211 \\
 &  2222111122203330 \\
 &  2000000000333222 \\
 \hline
 &  2210033222222222 \\
 &  2222221113333322 \\
 &  3333222233310001 \\
 &  3111111111000333 \\
 \hline
 &  3321100333333333 \\
 &  3333332220000033 \\
 &  0000333300021112 \\
 &  0222222222111000 \\
\end{array}.
$$
\hfill\quad $\blacksquare $
\end{example}

The next step is to construct a maximum period STGC for words whose length is $m^t$, where $t > 1$. Such a recursive construction
will be given when $m$ is an odd prime. For such an arrangement, it is required that the arrangement described in
Theorem~\ref{thm:self_dualST_NB} will have two additional properties. For each $i$, $1 \leq i \leq n$, there are two consecutive
words in the list that differ in position $i$. Such a requirement can be guaranteed either by seeds for a recursive construction
or by using the recursion. The second requirement is that the first sequence and the last one will also differ in $m$ coordinates.
The recursive construction has several steps, and it is a modification for the construction of binary STGC
of length $2^t$ and period $2^{2^t} - 2^{t+1}$ as was described in~\cite{EtPa96}.

For example we consider $m=3$, and construct the SDS $[V , V+\bfone , V+\bftwo]$ of period $3^{n+1}$, from the SDS $[X , X+\bfone , X+\bftwo]$,
where $V=(Z , Z+Y , Z+2Y+X)$, as proved in Lemma~\ref{lem:SDS_Z3}. The words $Z$ and $Y$ are of length $3^{n-1}$, where $Z$ is any word
over $Z_3$ that starts with a \emph{zero} and $Y$ is any word of length $3^{n-1}$ over $\Z_3$.

When $Z$ and $Y$ are fixed, then for $X$ we are using the component $X$ from $[X , X+\bfone , X+\bftwo]$ of all the
sequences of the arrangementofor length $3^n$
and period $3^{3^n}$. The outcome is a component yielding an arrangement
with $3^{3^n-n-1}$ SDSs of length $3^{n+1}$.

The next step is more complicated, and in this step, we merge all the components into one maximum period STGC of length $3^{n+1}$
and period $3^{3^{n+1}}$. For two components which share the same $Z$ and differ in position $i$ modulo $n$ of $Y$, the $X$'s used
will also differ in the $i$-th position in a way that the only part of $V$ in which the two SDSs differ is in one coordinate
(the $i$-th coordinate) of $Z+Y$.

If two components differ in the $i$-th position of $Z$, then the chosen $Y$'s will differ in the same coordinate in such a way
that both sequences will have the same values in $Z+Y$. The chosen $X$'s might be the same or also differ in the $i$-th coordinate such
that they will have the same value in $Z+2Y+X$ and they will only differ in one position in $Z$ (in the part $V=(Z , Z+Y , Z+2Y+X)$).

The last step is to guarantee in this recursive construction that for each $i$, $1 \leq i \leq n$, there are two consecutive
words in the list that differ in position $i$. Details
will be presented in the full version of this paper.

The same procedure will work for any given prime $p$, but merging the components will become more evolved. For example,
when $p=5$ we will have to consider $Z$, $Y_1$, $Y_2$, and $Y_3$ (see Section~\ref{sec:NB_SD}).
Having the right seeds for $p=3$ (with 729 SDSs of length 27 since the three of length 9 cannot satisfy the additional requirements)
and $p=5$, an STGC of length $p^t$ and period $p^{p^t}$ was constructed for each $t \geq 1$.

\vspace{-0.1cm}

\section{Conclusion}
\label{sec:conclusion}

Properties and recursive constructions of binary and non-binary SDSs were considered.
Connection between these sequences and maximum period STGCs
was described, and some guideline for the construction of such codes is given.
The first nontrivial families of maximum period STGCs were presented.

\noindent
{\bf Acknowledgment} -- The research was supported in part by the Israel Science Foundation under Grant 2462/24.

%

\end{document}